\documentclass[
showpacs,
onecolumn,floatfix,superscriptaddress,nofootinbib]{revtex4}
\usepackage{times,amsmath,amsfonts,amssymb,latexsym}
\usepackage{graphicx,epsf,epstopdf}
\usepackage[usenames,dvipsnames]{color}
\newcommand{\bra}[1]{\langle #1 |}
\newcommand{\ket}[1]{| #1 \rangle}

\newcommand{\be}{\begin{equation}}
\newcommand{\ee}{\end{equation}}
\newcommand{\ba}{\begin{eqnarray}}
\newcommand{\ea}{\end{eqnarray}}
\newcommand\tr{{\mbox{Tr\,}}}
\newcommand{\ignore}[1]{}
\def\CC{{\rm\kern.24em \vrule width.04em height1.46ex depth-.07ex
    \kern-.30em C}}
\def\P{{\rm I\kern-.25em P}}
\def\RR{{\rm
         \vrule width.04em height1.58ex depth-.0ex
         \kern-.04em R}}

\def\bbbc{{\mathchoice {\setbox0=\hbox{$\displaystyle\rm C$}\hbox{\hbox
to0pt{\kern0.4\wd0\vrule height0.9\ht0\hss}\box0}}
{\setbox0=\hbox{$\textstyle\rm C$}\hbox{\hbox
to0pt{\kern0.4\wd0\vrule height0.9\ht0\hss}\box0}}
{\setbox0=\hbox{$\scriptstyle\rm C$}\hbox{\hbox
to0pt{\kern0.4\wd0\vrule height0.9\ht0\hss}\box0}}
{\setbox0=\hbox{$\scriptscriptstyle\rm C$}\hbox{\hbox
to0pt{\kern0.4\wd0\vrule height0.9\ht0\hss}\box0}}}}
\def\bbbz{{\mathchoice {\hbox{$\sf\textstyle Z\kern-0.4em Z$}}
{\hbox{$\sf\textstyle Z\kern-0.4em Z$}}
{\hbox{$\sf\scriptstyle Z\kern-0.3em Z$}}
{\hbox{$\sf\scriptscriptstyle Z\kern-0.2em Z$}}}}

\begin{document}

\title{Background independent condensed matter models for quantum gravity}
\author{Alioscia Hamma}
\affiliation{Perimeter Institute for Theoretical Physics,
31 Caroline St. N, N2L 2Y5, Waterloo ON, Canada}
\author{Fotini Markopoulou}
\affiliation{Perimeter Institute for Theoretical Physics,
31 Caroline St. N, N2L 2Y5, Waterloo ON, Canada}
\affiliation{Department of Physics, University of Waterloo, Waterloo, Ontario N2L 3G1, Canada}
\affiliation{Max Planck Institute for Gravitational Physics (Albert Einstein Institute), 
Am M\"uhlenberg 1, D-14476 Golm, Germany.}
\affiliation{Santa Fe Institute, 1399 Hyde Park Road, 87501 Santa Fe, USA}

\begin{abstract}
A number of recent proposals for a quantum theory of gravity are based on the idea that spacetime geometry and gravity are derivative concepts and only apply at an approximate level.  There are two fundamental challenges to any such approach. At the conceptual level, there is a clash between the ``timelessness'' of general relativity and emergence.  Second, the lack of a fundamental spacetime makes difficult the straightforward application of well-known methods of statistical physics to the problem.  We recently initiated a study of such problems using spin systems based on evolution of quantum networks  with no a priori geometric notions as models for emergent geometry and gravity.  

In this article we review two such models.  The first is a model of emergent (flat) space and matter and we show how to use methods from quantum information theory to derive features such as speed of light from a non-geometric quantum system.  The second model exhibits interacting matter and geometry, with the geometry defined by the behavior of matter.  This model has primitive notions of gravitational attraction which we illustrate with a toy black hole, and exhibits entanglement between matter and geometry and thermalization of the quantum geometry.
\end{abstract}

 \maketitle
\section{Introduction}

Research in quantum gravity is currently at a very interesting point.  
In the past decade, the field has been presented with a great opportunity and challenge:  
quantum gravity may have observational consequences.  
In place of the long-standing frustration, that quantum effects of the gravitational field become important only at 
$10^{-35}m$, a tiny number we believed to be beyond observations, we now have investigations of
different sources of observational data, ranging from astrophysical (where large distances may amplify the tiny quantum gravity signals) to observational cosmology (where the CMB spectrum may carry signs of quantum gravity effects) to the LHC (for a recent review of quantum gravity phenomenology, see \cite{Hos}).  In all these cases, the theoretical explanation of the observations may require new physics involving quantum gravity.  It is exciting and challenging to find a theory that connects to these observations.

In this new environment, the novel direction of {emergent gravity} is particularly relevant.  
The central idea  is that spacetime geometry and gravity are derivative concepts and only apply at an approximate level. Quantum gravity is  one more situation in physics where we have the low energy theory -- general relativity and quantum field theory -- and are looking for the high energy, microscopic one.  As an analogy,  consider the example of 
 going from thermodynamics to the kinetic theory.  What we currently know is the low energy theory, the analogue of fluid dynamics.  We are looking for the microscopic theory, the analogue of the quantum molecular dynamics.   Just as there are no waves in the molecular theory,  we may not find geometric degrees of freedom in the fundamental theory.  
 Not surprisingly, this significant shift in perspective opens up new routes that may take us out of the old problems.

Results in this direction include
long-standing topics such as black hole thermodynamics \cite{Str} and matrix models \cite{MM}, the AdS/CFT correspondence and the appearance of gravitons in string theory, Horava gravity \cite{Hor}, emergence of a Lorentzian metric and aspects of gravitation such as Hawking radiation in analog models of gravity \cite{analog}, and condensed matter approaches 
such as G. Volovik's work on emergent Lorentz invariance at the Fermi point \cite{GV} and X.-G.\ Wen's work on emergent matter and gravitons from a bosonic spin system \cite{Wen}, the broader idea of spacetime as a condensate \cite{Hu}, or aspects of emergence in Causal Dynamical Triangulations \cite{CDTSO} and Group Field Theory, and similar ideas in a simplicial \cite{slee} and causal models \cite{dariano}.
 While approaches of this kind are newer and therefore less established,  this is a very promising and timely direction, especially in the light of the opportunities and challenges to quantum gravity research coming from the recent developments in observations.  
 At the same time, results in emergent gravity are mainly a collection of hints and analogies.  We are still lacking a microscopic theory that unifies and provides context to the existing results and, of course, that we can test against experiment.

We are particularly interested in what the emergence paradigm has to say about the problem of time.  
``Problem of time'' refers to the clash in the roles that time plays in General Relativity and quantum theory, and is considered fundamental in quantum gravity research since many other problems can be traced to it.  We can review it very briefly as follows (for classic accounts of the issue with far more depth, see, for example \cite{time}).

In General Relativity, spacetime, the 4-dimensional curved manifold with metric $g_{\mu\nu}$ and curvature $R_{\mu\nu}$, and matter,  the stress-energy tensor $T_{\mu\nu}$, are dynamical and affect each other:  matter tells spacetime how to curve and spacetime tells matter where to go.  This is the content of the Einstein equations,
\begin{equation}
R_{\mu\nu}-\frac{1}{2}g_{\mu\nu}R=\frac{8\pi G}{c^4} T_{\mu\nu},
\label{eq:EE}
\end{equation}
where $G$ is Newton's constant and $c$ is the speed of light.   
The Einstein equations are invariant under diffeomorphisms of the spacetime manifold, operations that map spacetime points to other spacetime points.  This means that spacetime coordinates are not physical, instead,  events and their relations are physical. A given collection of events and the order in which they occurred is physically meaningful but it can be represented by several different metrics, all related by diffeomorphisms.   The correct physical quantity is the equivalence class of metrics under diffeomorphisms, usually called a {\em geometry}.
Now let us consider a pure gravity scenario, a universe with no matter, where the right hand side of equation (\ref{eq:EE}) is zero. 
With no physical objects to  mark events, 
it is very difficult to construct observables that measure local physical properties. 
  In the 3+1 canonical 
 decomposition of the Einstein action, evolution is a diffeomorphism and it is impossible to locally disentangle the effects of change from the effects of changing coordinates.  The Hamiltonian for the evolution of space is just a constraint. 
This is what is often called the {\em timelessness} of General Relativity.

Timelessness becomes significant in quantum gravity. Quantum gravity is often seen as the problem of unifying or reconciling general relativity and quantum theory, combining the physics of the very large with that of the very small.   
Quantum theory always uses a fixed spacetime, and that time is a parameter, not even an operator.  
The incompatibility of general relativity and quantum theory can be stated in many ways and a classic one is {\em the problem of time}:  adapting the Scr\"{o}dinger equation to a diffeomorphism invariant context by quantizing  equation (\ref{eq:EE}) gives the Wheeler-deWitt equation,
\begin{equation}
\widehat{H}|\Psi_{\rm U}\rangle=0.
\label{eq:wdw}
\end{equation}
$|\Psi_{\rm U}\rangle$ is the wavefunction of the universe, $\widehat{H}$ is the quantum Hamiltonian constraint and 0 means there is no time.  

We believe that an emergence scenario for gravity can open up new potential solutions to the problem of time, mainly in two ways:
\begin{enumerate}
\item
{\em No pure gravity.}\/ 
Timelessness in the argument above arises in a universe of pure gravity.  If we allow matter, it is possible to define clocks and time.  While we do live in a universe with matter, general relativity allows for pure gravity solutions and there is no way to exclude them in the current setup.  Emergence of gravity may provide a mechanism for excluding pure gravity solutions, for example, if gravity is simply the thermodynamics of matter and  there is no gravity without matter.
\item 
{\em Fundamental vs geometric time.}\/  If gravity is emergent, then geometry and diffeomorphisms should also be  emergent.  It is then possible that one can have a fundamental time, consistent with quantum mechanics, at the microscopic level, and an emergent or geometric time, the $g_{00}$ component of the emergent metric, macroscopically.  
Quantum theory and General Relativity could be reconciled if we have a reason why they should apply at different levels, and emergence can provide that reason\footnote{
An appealing, for us, feature of this scenario is that if diffeomorphism and Lorentz invariance are emergent symmetrues, there will be departures from the exact symmetry, providing observational tests for theories of emergent gravity.
}.
\end{enumerate}

 Our aim is twofold.  First, we wish to study emergent gravity from the perspective of background independence and time in quantum gravity.  In addition, we would like to see a unified microscopic picture behind all the various hints of why gravity may be emergent.   Emergence for us means that as we go to the Planck scale, the Hamiltonian model becomes very complicated and messy. It is at low energies that symmetries and the simplicity of physical laws emerge. This kind of emergence is very anti-reductionist because the building blocks of Nature can be very complicated (and not fundamental, they perhaps are made of even more complicated building blocks), whereas the emergent structures are simple and beautiful.  In the case of gravity, this means, for example, that diffeomorphisms should appear as an emergent symmetry.
 
 We will take a straightforward approach to the problem.  
 Emergence is studied primarily in statistical/condensed matter physics.  The classic paradigm is the Ising model: the microscopic physics is described by spins on a lattice, while the emergent physics in the right phase can exhibit ferromagnetism.  Ferromagnetism is an emergent phenomenon, as it is not a property of the microscopic degrees of freedom, instead, it is an ordering of the collective.  
The possibility that gravity may be emergent suggests that quantum gravity ought to be studied as a problem in statistical physics.  
The question we will be asking is {\em ``What 
 is the analogue of the Ising model for gravity?''} 
 
In this direction, we propose to study {\em background independent spin systems} to model emergence of geometry and gravity.  We will attempt to adapt the methods and pradigms of statistical physics to the analogue of dynamical geometry: we will study spin systems on a {dynamical lattice}.  We initiated this program in \cite{KoMaSm,KoMaSe}, where we proposed 
 {\em quantum graphity}, a spin system on a dynamical lattice whose purpose was to study emergence of flat space and matter, and developed this further in \cite{HMLCSM} with another such spin system, modeling interaction of geometry and matter.   While this is a new direction of research, we have
 studied several subjects, including the emergence of (flat) geometry and matter \cite{KoMaSm,KoMaSe}, deriving the speed of light from first principles \cite{hlr}, matter/geometry interaction and entanglement and issues in quantum cosmology \cite{HMLCSM}, and the distinction between fundamental and emergent time \cite{FQXi}.
 Not surprisingly, we find that the change in perspective that emergent gravity brings does indeed raise new opportunities.  It also creates new puzzles, which we hope to be able to clarify in the relatively straightforward context of spin systems.   The purpose of this contribution is to review the models and results and outline future directions of research and the major open issues.  
 
 The outline of this article is as follows. 
In Section \ref{basicidea}, we describe the basic idea of graphity models, which is to describe the dynamical lattice using the space of adjacency qubits of the complete graph.  We explain what we mean by using a graph as space, and outline the task involved in finding the right Hamiltonian for the spin system.  In Section \ref{Model1}, we review the first quantum graphity model of \cite{KoMaSe}.  The Hamiltonian for this model is an extension of the string network condensation mechanism of Wen and collaborators for emergent matter, which we review before extending it to dynamical lattices.  We then describe a method to derive the emergent speed of light from the microscopic physics using the Lieb-Robinson bound for information propagation in a local spin system and discuss the extend to which this means that the model has an effective Lorentzian spacetime.  We end this section by presenting a new tool for dealing with dynamical graphs, the transformation from the complete to the line graph.  In Section \ref{Model2}, we review the model of interacting matter and geometry that was proposed in\cite{HMLCSM}.  In this model, we have attempted to implement the idea of ``geometry tells matter where to go and matter tells geometry how to curve'' in a spin Hamiltonian.  The question is to what extend this captures elements of gravitational behavior and we study it by looking at a toy black hole.  In addition, the model produces entanglement between matter and geometry, which gives rise to the possibility that each of the two, studied separately, will show thermalization.  We present results that support this.  In Section \ref{Future}, we outline the major issues and directions for this program of research, namely, the quantum behavior of the matter/geometry interaction, time and Lorentz invariance, and gravity.

\section{Spin systems on a dynamical lattice}
\label{basicidea}

Adjacency, an essential aspect of geometry, is fixed in spin systems such as the Ising model.  The spin system is given by a Hamiltonian on a lattice and this lattice determines adjacency.  A local Hamiltonian then is a sum of terms of neighbor interactions.  For example, in the Ising model, the Hamiltonian is $H=-\sum_{\langle ij\rangle} J_{ij}\sigma_i\sigma_j$, where the sum is over all adjacent spins $i$ and $j$.  As explained above, we instead want adjacency to be dynamical and
so we will turn adjacency into a quantum degree of freedom.  

We do this by starting with the lattice of {\em all possible} adjacencies.  For $N$ systems, this is $K_N$, the complete graph on $N$ vertices.  
$K_N$ has $N\choose 2$ links, corresponding to all possible pairings of its $N$ vertices.  
  For example, for $N=4$, the graph of all possible pairs $ij$ with $i,j=1,...,4$ is $K_4$:
\be
	\begin{array}{c}\mbox{\includegraphics[height=1.5cm]{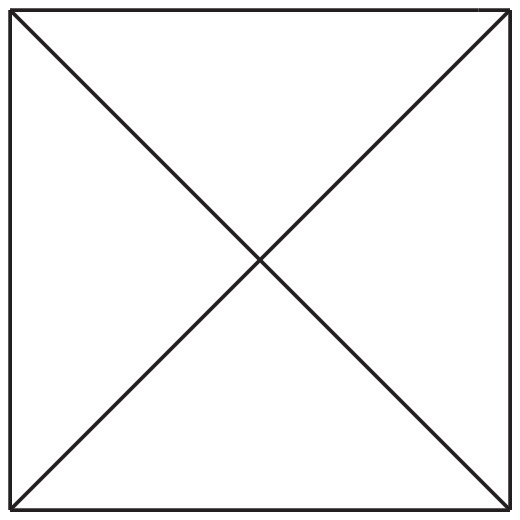}}\end{array}\nonumber
\ee
We now give each link  $e_{ij}$ in $K_N$ a Hilbert space ${\cal H}_{e_{ij}}\simeq{\bf C}^2$ with basis elements $\{|1\rangle,|0\rangle\}$,  and interpret $|1\rangle_{ij}$ to mean that the edge $e_{ij}$ is {\em on} and $i,j$ are adjacent, and $|0\rangle$ to mean the edge is {\em off}.  
The total state space then is
\be
{\cal H}_{\rm graph}=\bigotimes_{e=1}^{\frac{N(N-1)}{2}}{\cal H}_e.
\label{eq:Hgraph}
\ee
By having {\em on} and {\em off} degrees of freedom on the edges, a generic state in ${\cal H}_{\rm graph}$ is a superposition of subgraphs of $K_N$, which for large enough $N$ and to the extend that graphs can represent geometry, can be thought of as quantum superposition of quantum geometries.  


\subsection{ Graphs and space.}

We will utilize the graph of adjacencies as a primitive form of geometry.  Adjacency of course defines neighborhoods, a basic aspect of topology.  Together with a local Hamiltonian it determines a causal set of events, or interactions.  A causal set has long been known to contain the information of the metric, up to the conformal factor.  

Graphs are used in many different approaches to quantum gravity, from spin networks in quantum gravity \cite{Pen,HasPer} and Loop Quantum Gravity \cite{snets}, Spin Foams \cite{SF} and Group Field Theory \cite{GFT}, to the partial orders in Causal  Sets \cite{CS} and Quantum Causal Histories \cite{QCH}.  There is substantial freedom in the assignment of geometry to a graph.  One can consider the graph as the dual to a piecewise linear triangulation of some given dimension, or use measures of geometry such as the Hausdorff dimension of the graph.  A unified and organized way of assigning a geometry to a general graph does not currently exist in quantum gravity research.  For example, even if the evolution of our states defines a (quantum) causal set, recovering the metric when one only has the causal set is a formidable problem.

Fortunately for our purposes, the correspondence between a graph and space is clear in two extremal situations:  no space and flat, or nearly flat, space.  $K_N$ represents no space.  The neighborhood of  a vertex in $K_N$ is the entire $K_N$, meaning that in that state of adjacency there is no notion of subsystems, and hence no notion of locality.  We interpret this state as no space.  On the other hand, if the graph is a regular lattice, say, a cubic lattice, there is also a straightforward interpretation of it as flat space in 3 dimensions.  

Back to the state space ${\cal H}_{\rm graph}$, we note that, for very large $N$, the regular cubic lattice is a subgraph of $K_N$, which means a particular state in ${\cal H}_{\rm graph}$.  
Simply using these two extremal cases we can sketch out the basic idea behind our models.  We will think of {\em space as order}, and will consider space to be emergent when geometrical symmetries such as the homogeneity and isotropy of the FRW metric appear.  $K_N$ in contrast is permutation invariant.  The scenario of {\em geometrogenesis} proposed in \cite{KoMaSm,KoMaSe} is that at early times, or high energies, the universe is in the $K_N$ phase, where there is no useful notion of space, and at low energy it freezes in the symmetric state of an FRW metric.  The questions regarding emergence of geometry will then be: A generic state in ${\cal H}_{\rm graph}$ is a quantum superposition of graphs and, therefore, to the extend that the graph represents a geometry, a superposition of geometries.  We do not see such superpositions macroscopically.  Why?  In addition, can the system settle in a regular lattice as its ground state or long-lived metastable state?


\subsection{Purpose of models.}
The task in constructing quantum graphity models is to find a Hamiltonian such that:
\begin{enumerate}
\item
Shows how a regular, smooth geometry emerges from disordered or {\em no} geometry.
\item
Exhibits primitive notions of gravity, e.g., attraction, horizons, negative heat capacity, etc, at the collective level.
\item
Lets us investigate quantum effects such as quantum interference of geometries and entanglement of matter and geometry.
\item
The model can be used to investigate questions in emergent gravity, such as time vs emergence, and develop new tools for emergent gravity physics in an explicit context.
\end{enumerate}

In this direction, we have already studied several subjects, including 
mergence of (flat) geometry and matter \cite{KoMaSm,KoMaSe}, deriving the speed of light from first principles \cite{hlr}, matter/geometry interaction and entanglement and issues in quantum cosmology \cite{HMLCSM}, as well as fundamental vs geometric time \cite{FQXi}. 

We will now review the two models of \cite{KoMaSm,KoMaSe} and \cite{HMLCSM}.

\section{Quantum graphity: A model of emergent geometry and matter}
\label{Model1}

{\em Quantum graphity}, first proposed in \cite{KoMaSm} and \cite{KoMaSe}, is a spin system in which locality, geometric symmetries, and matter arise at low energy, when a spin system is cooled to its ground state.  The model is an extension of the string-net condensation mechanism for emergent $U(1)$ matter, photons and fermions proposed by X.-G.\ Wen and collaborators\cite{Wen}.  This model is on a fixed lattice and we extended it to the dynamical quantum lattice defined by eq.\ (\ref{eq:Hgraph}).  The result is that the microscopic degrees of freedom that determine the ground state lattice are the same as those that give rise to the U(1) matter in the appropriate limit.

\begin{figure}
  \includegraphics[width=8cm]{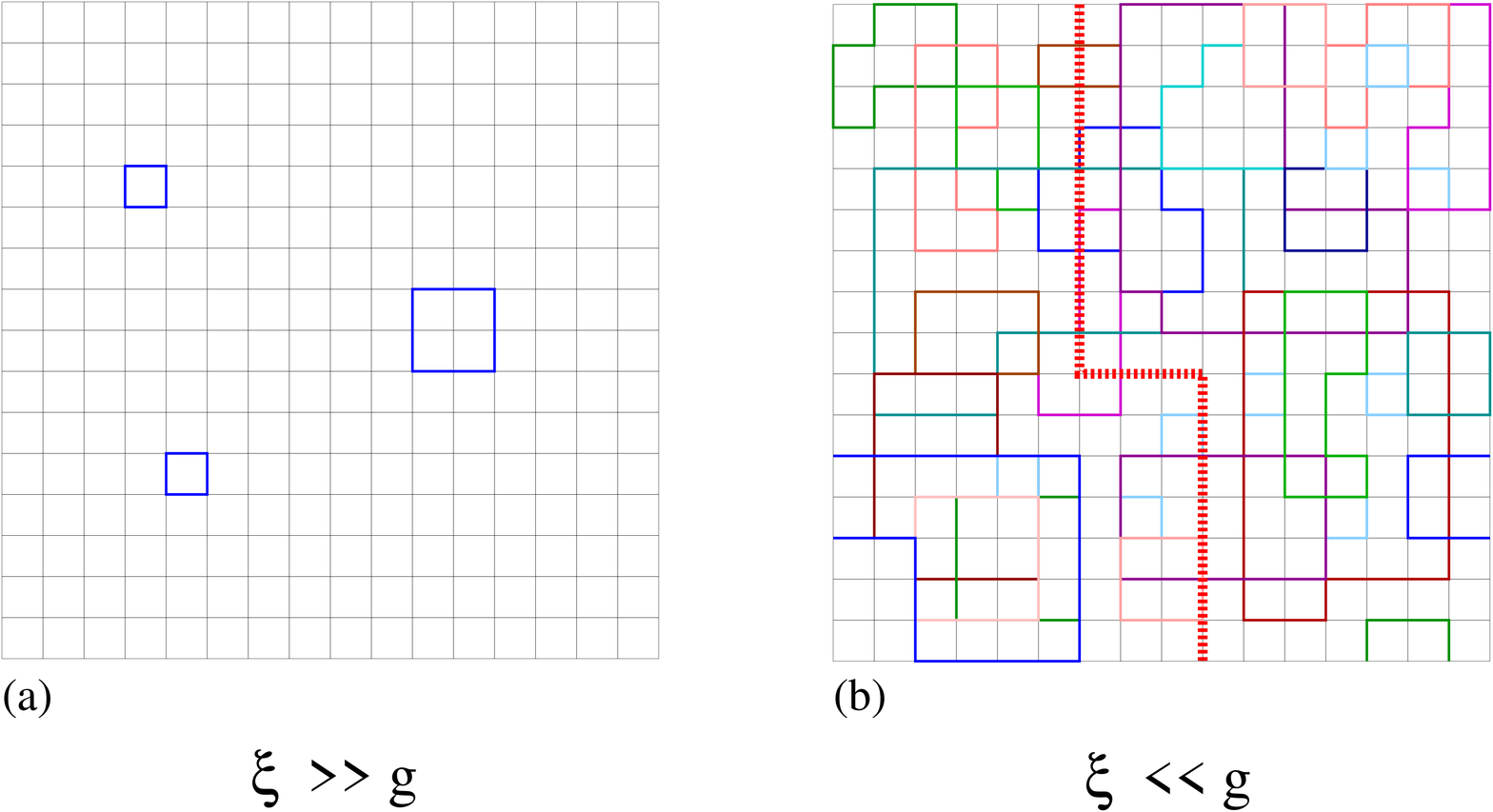}\\
  \caption{Two configurations of colored strings, or string-nets: (a) The black links are in the state $|1,0\rangle$ and the blue links in the state $|1,1\rangle$, while all the other edges in $K_N$ are in the state $|0,0\rangle$. (b)
A state with many colors. Again, all the edges in $K_N$ that are not drawn are in the state $|0,0\rangle$. In both figures the 
$\{ j_{ab}\}$ degrees of freedom are frozen in the configuration of a square lattice. }
  \label{stringnet}
\end{figure}

The state space of this model is  ${\cal H}_{\rm graph}=\bigotimes_e{\cal H}_e$, with the edge labels extended to 
\be
{\cal H}_e={\rm span}\left\{|j_{ij}, m_{ij}\rangle\right\}=
{\rm span}\left\{
|0,0\rangle, |1,-1\rangle, |1,0\rangle,|1,1\rangle\right\}\simeq{\bf C}^4.
\ee
As previously, $j_{ij}=\{|1\rangle,|0\rangle\}$ means that the edge $e_{ij}$ is {\em on} or {\em off} respectively.  With the $m_{ij}$ labels, we have one {\em off} state and 3 {\em on} states. A finite dimensional Hilbert space at every edge means that this is nothing else than a spin system. In our interpretation, a configuration of {\em on} states is a graph $\Gamma_N$. If the graph has enough order, it will represent space. Now we see that an {\em on} edge can have three different values, or colors. These colors allow us to draw networks on $\Gamma_N$. For instance, if $\Gamma_N$ is a square lattice, a basis state in $\mathcal H_{\rm graph}$ can now be a configuration of colored strings, or string-nets, like in Fig.\ref{stringnet}. 
Following the mechanism of string-net condensation invented by X.-G. Wen and collaborators \cite{Wen},  the strings of colored edges will give rise to an emergent gauge field or field of matter. A string condensate is a a liquid of fluctuating strings, with the collective excitations above such states corresponding to gauge bosons and fermions. In string net condensation, just as phonons are emergent collective excitations of a crystal, gauge bosons and fermions are collective excitations of a pattern of strings.  We will outline this mechanism below.

Let us first define the following four operators acting on each ${\cal H}_e$,
\be
\begin{array}{rl}
J|j,m\rangle&=j|j,m\rangle\\
M|j,m\rangle&=m|j,m\rangle,
\end{array}
\ee
and the raising and lowering operators 
\be
\begin{array}{rl}
M^+|j,m\rangle&=
\frac{1}{\sqrt{2}}\sqrt{(j-m)(j+m+1)}|j,m+1\rangle\\
M^-|j,m\rangle&=
\frac{1}{\sqrt{2}}\sqrt{(j=m)(j-m+1)}|j,m-1\rangle.
\end{array}
\ee
The relevant commutation relations are
\be
\left[M^+,M^-\right]=M,\quad
\left[M,M^\pm\right]=\pm M^\pm.
\ee
It is important that all these operators annihilate the {\em off} state:
\be
J|0,0\rangle=
M|0,0\rangle=
M^\pm|0,0\rangle=
0.
\ee

Now, on each edge of $K_N$ we can define creation and annihilation operators $a^\dagger$ and $a$, with $a|0\rangle=0, a|1\rangle=|0\rangle$, and the number operator $n$, $a^\dagger a|n\rangle=|n\rangle$.  We use these to define the operator $N$, a quantum analogue of the adjacency matrix of a graph: $N_{ij}$ is the number operator acting on edge $e_{ij}$ in the graph and returning the value of that edge, e.g.,
$
N_{13}\left(|e_{12}\rangle\otimes|e_{13}\rangle\otimes...\right)=n_{13}
	\left(|e_{12}\rangle\otimes|e_{13}\rangle\otimes...\right)
$.
Powers of $N_{ij}^{(L)}$ tell us how many paths of length $L$ connect $i$ and $j$. 
For example, paths of length 2 are counted by 
$
N^{(2)}_{ij}=\sum_k N_{ik} N_{kj}$, etc.
The important property of these operators that we will use in this model is that normal-ordering $N$ returns the number of {\em non-overlapping} paths from $i$ to $j$  of length $L$ (for why this is so, see \cite{KoMaSe}).  

We can now define the Hamiltonian of the model.  It has the following terms:
\begin{itemize}
\item
Vertex term: 
\be
H_V=g_V \sum_i e^{\left(v_0-\sum_j N_{ij}\right)^2},
\ee
for $g_V>0$ and $v_0$ a free parameter.  This term is minimized when all vertices have degree $v_0$.
\item
Loop (and string) terms:
Using the full state space (the $m$-values on the edge Hilbert spaces), the Hamiltonian that determines the loop distribution in the ground state graph is a generalization of Wen's string-net condensation \cite{Wen} to a dynamical lattice.  
A string condensate is defined by the types of strings of the model, namely the ``colors'' we use, and the way we allow them to branch. 
To explain the mechanism, let
us start by freezing all the edges in some  configuration $\Gamma_N$. Then we can consider a Hamiltonian that is the sum of a constraint term, a potential energy term, and a fluctuation term:
\be\label{terms}
H_{\rm strings} = V H_C + \xi H_{\rm pot} + g H_{\rm kin}.
\ee
The constraint term says that it costs a large energy $V\gg\xi,g$ to have open strings. So, at low energies, all the allowed string configurations are closed string-nets, like in Fig.\ref{stringnet}(a-b). The potential term $\xi$ is a string tension. When $\xi\gg g$ the energy of a string configuration is proportional to its length and therefore only a few small strings are allowed in the ground state, see Fig.\ref{stringnet}(a). On the other hand, when $\xi \ll g$ the kinetic energy makes the string fluctuate a lot and the ground state is a superposition of many closed string configurations with arbitrary length, as in Fig.\ref{stringnet}(b). This is the so-called {\em string condensed state}. The modes of vibration of these strings are the first category of excitations above the ground state. The second type is made of defects, namely violations of the constraint $V H_C$. In the string pictures, such defects are endpoints of open strings. In \cite{Wen}, it was shown that the fluctuations of closed strings correspond to $U(1)$ gauge bosons (photons) while endpoints of open strings correspond to charged fermions  (electrons). 

To extend this construction to the case in which the $\{ j_{ij} \}$ degrees of freedom are not frozen in a particular lattice $\Gamma_N$ like the square lattice of Fig.\ref{stringnet}(a-b),  in \cite{KoMaSm,KoMaSe} we modified the Hamiltonian of \cite{Wen} to the terms:
\be
H_{\rm loops}
	=-g_L\sum_{\rm loops}\sum_{L=0}^\infty\frac{r^L}{L!} \prod_{a=1}^L M_a^\pm,
\label{eq:Hloops}
\ee
with $\prod_{a=1}^L M_a^\pm=M_{ij}^+ M_{jk}^- ... M_{yz}^+ M_{zi}^-$, 
and
\be
H_{\rm strings}=g\sum_i\left(\sum_j M_{ij}\right)^2+ \xi\sum_{ij}M_{ij}^2.
\ee
$g$ and $\xi$ are positive couplings.  
The ground state of $_{\rm string} $ consists of all links having $m=0$.  Again, the $g$ term is the kinetic energy term that makes strings fluctuate while the $\xi$ term is a potential energy corresponding to a string tension. 

At the same time, the term $H_{\rm loops}$ also determines the ground state lattice.  This can be seen more easily if we consider the term eq.\ (\ref{eq:Hloops}) acting on the reduced state space ${\cal H}_{|0,1\rangle}$, i.e., ignoring the $m$ labels.  In that case, the term (\ref{eq:Hloops}) can be written as
\be
H_{\rm loops}=-g_L\sum_{i,j}\delta_{ij}\ e^{r N_{ab\\ij}}
	=-g_L\sum_{i,j}\delta_{ij}\sum_{L=0}^\infty\frac{r^L}{L!} N_{ij}^{(L)}.
\ee
When normal ordered, $:H_{\rm loops}$ counts all non-overlapping paths of different lengths and lowers the energy when the number of loops of length $L^*$ is maximized, where $L^*$ depends on the choice of the parameter $r$.  In \cite{KoMaSe}, we found a choice of parameters ($v_0=3$ and $r\ge 7.1$) for which  {\em the honeycomb lattice is a stable local minimum}. 

\item
Evolution terms:
The terms above are eigen-operators of graph states and do not change the linking structure between vertices.  The time evolution of the model should include terms that change the graph.  In \cite{KoMaSe}, we suggested local graph evolution moves, much like one does in spin foam models, for example by terms that exchange linking or add/subtract edges:
\be
	\begin{array}{c}\mbox{\includegraphics[height=1.5cm]{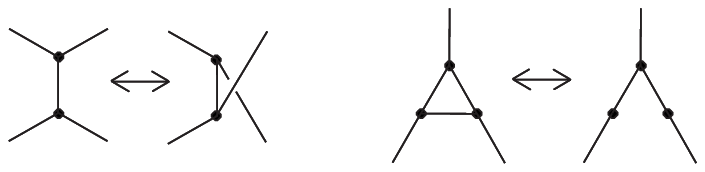}}\end{array}\nonumber
\ee
So far, the focus of our work in this model has been to look for the lowest energy graph states, for which this part of the Hamiltonian is not required.  As a result, we have no  insight on the particular properties of some type of evolution over another.   
\end{itemize}
 
The full Hamiltonian of the model,
\be
H=H_V+H_{\rm loops}+H_{\rm strings},
\ee
 (normal ordered) is then a modification of the string-net condensation mechanism of \cite{Wen} in that the supporting lattice is dynamical.  The interesting feature of the model is that the same degrees of freedom give rise to the $U(1)$ gauge bosons and charged fermions and also shape the lattice.  This can be thought of as a novel form of unification of the matter and geometry, with the separation between the two appearing only at the emergent level.  
In the right region of the couplings, the model has a stable energy minimum that exhibits both a lattice configuration (the honeycomb) and string-net condensation \cite{KoMaSe,Kon}\footnote{
A similarly motivated model of emergent regular space was recently proposed in \cite{Con}.
}. The space, fields and electrons have emerged all together.

\subsection{Speed of light via the Lieb-Robinson bound.}
\label{SectionLR}

We have seen that in string-net condensation,
 in the limit $\xi \ll g$, one obtains emergent photons as collective excitations of fluctuating strings. In the continuum limit, these excitations are described by the Lagrangian of electromagnetism and the speed of light is a function of the microscopic couplings,  $c = \sqrt{\xi g}$. An important question now arises:
 Does the emergence of the Maxwell equations imply the emergence of Lorentz invariance? Or is the fact that the elementary excitations behave like photons not enough to have Lorentz invariance? Indeed, do we even have a light cone? What forbids other signals to travel faster than light? 

The system we are considering is not relativistic, after all, we are using ordinary non-relativistic quantum mechanics. It is well known that in principle in non-relativistic quantum mechanics it is possible to send signals arbitrarily far away in an arbitrarily short time. This is easy to understand in perturbation theory. If we perturb some eigenstate of a quantum many-body system with a local perturbation, for example by flipping one spin in a long spin chain, the system will be away from equilibrium and will start evolving in time. For some arbitrarily short time $t_\epsilon$, there is an order in perturbation theory such that a spin arbitrarily far from the perturbed spin will be away from equilibrium as well, and will feel the perturbation. We know that ordinary non-relativistic quantum mechanics is not the right theory and one should use quanutm field theory instead, still, it seems strange that such a violation of causality is possible. Well, the point is that such violation is not that strong after all. If the spins are very far away and the time is very short, to see the effect one must go to a \textit{very} high order in perturbation theory.   In effect, this defines an emergent lightcone, outside which the signal is exponentially supressed.  

The key for this to happen is locality. 
The type of system we are considering goes under the name of \textit{local bosonic system}. A local bosonic system is a system in which locality is constrained at the fundamental level. By this we mean that the Hilbert space is the tensor product of small local Hilbert spaces, and that the Hamiltonian is the sum of local operators with support on only a finite (and small) number of local Hilbert spaces. In this case, one can formalize the emergent lightcone structure in a rigorous way, using the Lieb-Robinson bound,  a bound to the violation of causality that one can have \cite{lr}. This bound implies the existence of a maximum speed $v_{\rm LR}$ of signals in the medium. The speed $v_{\rm LR}$ determines an effective light cone such that signals outside the light cone are exponentially suppressed. 

To be more precise, consider different-time commutators of observables with support on regions $P,Q$ of the system that are spatially separated by a distance $d_{PQ}$. Consider a Hamiltonian that is a sum of local terms $H=\sum_{\langle ab\rangle}h_{ab}$ with $\langle ab \rangle$ the edges of a graph. Then we can bound the commutator between two observables $O_P(t), O_Q(0)$ as 
\be
\| [O_P(t) , O_Q (0)] \| \le 2\|P_P\| \|O_Q\| C \exp \{a (d_{PQ} -v_{\rm LR}t)\},
\label{LR}
\ee
where the constants $a,C$ depend on the graph and the speed $v_{\rm LR}$ depends on both the graph and the maximum strength of the interactions $\| h_{ab} \|$ \cite{lr}.  In \cite{hlr}, we applied the Lieb-Robinson theorem to Wen's Hamiltonian for emergent photons to find the Lieb-Robinson speed of the emergent $U(1)$ excitations. Usually the problem with the bound is that it is too loose because it overkills by choosing the maximum strength of interactions. This would yield an estimate for the maximum speed of signals $v_{\rm LR} \simeq g$ which is much higher than the speed of emergent light $c$. We were able to find a tighter bound in which $v_{\rm LR}\simeq \sqrt{\xi g}\equiv c$ (recall that the speed of light in the emergent Maxwell equations is related to the fundamental couplings by $c=\sqrt{\xi g}$). Therefore emergent light is also the fastest signal in the system. 

This is a strong argument in favor of the emergence of Lorentz symmetry from a spin system.  However, more work needs to be done in this direction.  We also note that the above derivation of an emergent light cone does not address the problem of different species traveling at different speeds, a problem generic to emergent gravity approaches.

\subsection{Transition in the causal structure}
An important feature of the Lieb-Robinson speed defined by the bound is that it depends on the 
degree of the vertices on the lattice: 
\be 
v_{\rm LR}\propto d.
\ee
  One can understand this intuitively:  {the transmission of information and correlations through the lattice is a quantum effect due to the sum over all the possible paths that connect two points. Therefore, the higher the number of paths that connect two points $P$ and $Q$ on the lattice, the stronger the signal at $P$ from $Q$}.   Or, equivalently, more paths open up the lightcones and increase $v_{\rm LR}$.  The vertex  degree is a measure of the number of paths between two points in the lattice, which is why we find that the LR speed is proportional to it.  
 In section \ref{toyBH}, we will use this fact to describe a toy black hole.  

For the model defined above, $v_{\rm LR}\propto d$ means that at high energy, when more edges are {\rm on}, and therefore the average vertex degree is higher, the speed of light is also higher.  If, for example, we approximate the transition from high to low energy states by a sequence of hypercubes of dimension $D$, then, since $D\propto d$, we will find that $v_{\rm LR}\propto D$, and hence it drops with the dimension.  

In \cite{hlr}, we pointed out that this effect can be used to resolve the homogeneity problem in cosmology.  This is the question of why we see the sky at a uniform temperature, the cosmic microwave background, including distant parts that have had no causal connection in the past.  It is usually solved by assuming an inflationary period in the early universe.  Here, the causal mechanism for thermal equilibrium is the opening up of the lightcones caused by more {\em on} edges at higher energies (early times).  This is interesting as it illustrates the possibility of a transition in the causal structure and that such a quantum gravity effect can have observable consequences that are not limited to the Planck scale.  Of course, further work is needed to check if a variation of the speed of light like the one proposed here is consistent with ither  observations.  We note that, while we can always defined a Lieb-Robinson speed of information propagation, for a highly connected lattice the string network condensation mechanism does not apply, and we do not expect $v_{\rm LR}$ to be the speed of light as there is no emergent light.


\subsection{From a dynamical to a fixed lattice}

Central to our motivation for studying emergent gravity using spin systems is to gain access to  tools to describe emergence used in statistical physics and condensed matter theory.  However, to accommodate the possibility of dynamical geometry, we are using spin systems on dynamical lattices.  Many of the tools that we would like to use are of course designed for spin systems on a fixed lattice.  For example, we would like to use mean field theory analysis to find out if the transition to the flat graph described above is a phase transition.  How can we do that if the connectivity of the lattice is a variable? 

It turns out that the basic idea of writing the ensemble of graphs as subgraphs of $K_N$ for some large enough $N$ also allows us to transform the dynamical lattice in the model above to a spin system on a fixed lattice.  This can be done by mapping $K_N$ to its {\em line graph} $L(K_N)$, as follows \cite{CaMa}:
Every edge  of $K_N$ maps to a vertex in $L(K_N)$.  Thus, $L(K_N)$ is a graph with $N(N-1)/2$ vertices.  We now connect two vertices in $L(K_N)$ if the corresponding edges in $K_N$ shared a vertex.  For example, for $K_4$, the original graph and its line graph $L(K_4)$ are:
\be
	\begin{array}{c}\mbox{\includegraphics[height=1.5cm]{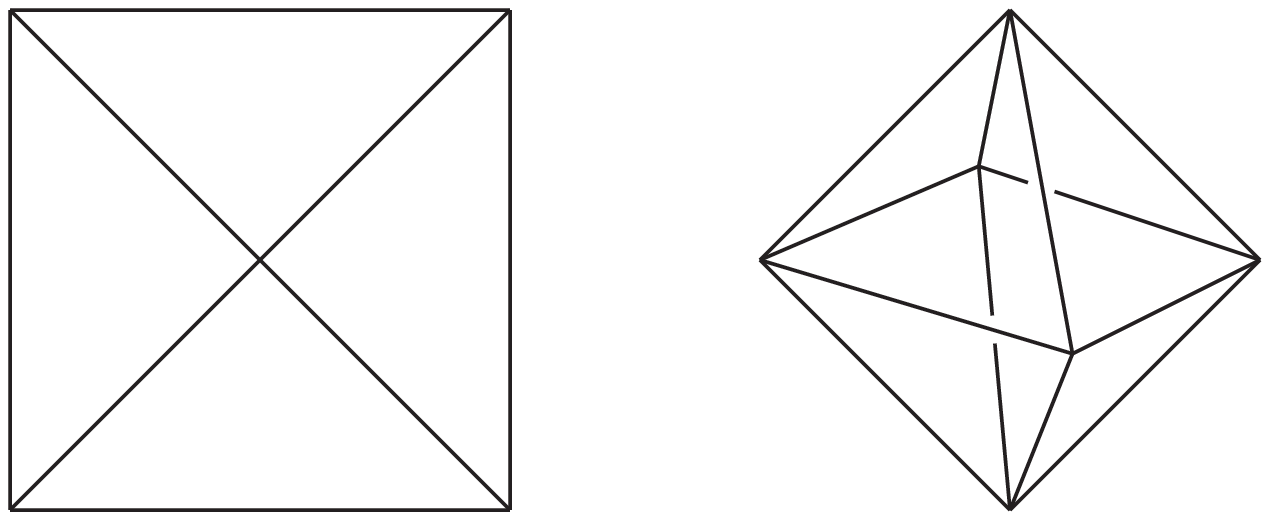}}\end{array}\nonumber
\ee
It is now easy to see that the values $j_e=\{|1\rangle, |0\rangle\}$ on the original system become simple quantum Ising spins on the vertices of the line graph.  A particular graph, a state  in ${\cal H}_{\rm graph}$, is simply an Ising spin configuration on the fixed lattice $L(K_N)$.  

In  
\cite{CaMa}, we used this mapping 
in a weak coupling approximation of the model. With a mean field theory approximation and  low temperature expansion, we studied the properties of the model near zero temperature. We found that the model is dual to an Ising model with external nonzero magnetic field if we neglect the interaction terms. In particular, we showed that the average vertex degree is naturally a good order parameter for the mean field theory approximation and we found that  the parameter $v_0$ plays the role of the external magnetic field. Since $v_0$ is  never zero, the model has no phase transition and at $T = 0$ the system goes to the ground state as expected.
We also confirmed that, as the temperature drops, the speed of the emergent light must drop  with the vertex degree.

This method of dealing with dynamical graphs is likely to be of more general interest in background independent quantum gravity approaches, where dynamical graphs and ensembles of graphs are commonly used.  As long as in the theory in question an isolated vertex  is physically equivalent to no vertex, the ensemble of graphs can always be written in terms of the state space ${\cal H}_{\rm graph}$, that is, the graphs can be viewed as subgraphs of $K_N$ for some sufficiently large $N$.  If the degrees of freedom reside on the edges only, then the above mapping  turns a quantum ensemble of graphs into an Ising-like model.  For theories with degrees of freedom on the vertices (as in general spin foam models for example), a generalization of the above mapping may be possible.

\section{A unitary model of interacting matter and geometry}
\label{Model2}

In the previous section we described a background independent model for emergent locality, spatial
geometry and matter.  This model leaves open several questions.  An important one is that
in the scenario
of \cite{KoMaSe}  the universe starts at a high temperature
configuration ($K_N$) and evolves to a low-energy one (a regular local lattice). This has clear
limitations when applied to a cosmological context. In effect, what we described above assumes an external heat bath in which we can dump the energy of all the links that we turn off.  What we would like instead is a unitary, energy-conserving model.  In \cite{HMLCSM}, we constructed such a model, 
 in which graph edges can be deleted and matter created
and vice versa. We can turn links off and still conserve energy because we can transfer the energy to from the links to the matter.  
 This  is essentially an extension of the Hubbard model
to a dynamical lattice.  
The second motivation for the model we are about to present now goes back to the interpretation of General Relativity as ``geometry tells matter where to go and matter tells geometry how to curve''.  We wish to know if it is possible to implement this kind of behavior in a spin system, and if so, study to what extend such a spin system can show aspects of gravitation.

The model of \cite{HMLCSM} then is motivated by the questions:
\begin{enumerate}
\item
If gravity is emergent, is there an analogue of 
the Ising model for gravity?  Can we get horizons, attraction, negative heat capacity, etc, from a spin system with a local Hamiltonian? 
\item 
The spin system ultimately should describe a cosmological theory (just as General Relativity does at the emergent level).   Can we have unitary microscopic evolution that equilibrates to appear approximately thermal for long enough times? 
\item
In a system with state space ${\cal H}_{\rm geometry}\otimes{\cal H}_{\rm matter}$, study quantum effects of the matter/geometry interaction, such as entanglement of matter and geometry and superposition of quantum geometries.
\end{enumerate}

We began addressing the above questions in \cite{HMLCSM} and in \cite{th}.  
Geometry in this model is described, as before, by basis states $\{|1\rangle,|0\rangle\}$ on the edges $e$ of complete graph $K_N$ on $N$ vertices, interpreted as the edges being on or off respectively. That is, ${\cal H}_{\rm graph}=\bigotimes_{e\in K_N}{\cal H}_e$; ${\cal H}_e\simeq{\bf C}^2$. 
We implemented quantum matter in the simplest way possible, by allowing for bosons that live on the vertices of the graph: ${\cal H}_{\rm matter}=\bigotimes_{i=1}^N{\cal H}_i$, where ${\cal H}_i$ is the state space of a harmonic oscillator on vertex $i\in K_N$
The state space of the model then is:
\be
{\cal H}=
{\cal H}_{\rm graph}\otimes{\cal H}_{\rm matter}=
\bigotimes_{e\in K_N}{\cal H}_e
\bigotimes_{i=1}^N{\cal H}_i, 
\ee
with basis states of the form
\be
|\Psi\rangle=|\Psi_{\rm graph}\rangle\otimes|\Psi_{\rm matter}\rangle=
|e_1,...,e_{\frac{N(N-1)}{2}}\rangle\otimes|n_1,...,n_N\rangle.
\ee
We give the links energy when they are in the {\em on} state:
\be
H_{\rm link}=-U\sum_{ij}\sigma_{ij}^z,
\ee
while the boson energy is given by
\be
H_v=\sum_{i=1}^N H_i=-\sum_i \mu_ib^\dagger_i b_i,
\ee
where $b_i, b_i^\dagger$ are the annihilation and creation operators for a boson on site $i$.  

As in the previous model, we can have  superposition of interactions, or graph edges.  For example, the state
\be
|\psi_{ij}\rangle=\frac
{|10\rangle\otimes|1\rangle_{ij}+|10\rangle\otimes|0\rangle_{ij}}
{\sqrt{2}}
\ee
describes a particle at $i$ and a particle at $j$ and a superposition of them interacting and not interacting, or, equivalently, a superposition of geometries.  The state
\be
|\phi_{ij}\rangle=\frac
{|00\rangle\otimes|1\rangle_{ij}+|11\rangle\otimes|0\rangle_{ij}}
{\sqrt{2}},
\ee
on the other hand, describes {\em entanglement of matter with geometry} and distinguishes this model from the previous one.

The basic idea in this model is that the energy of a graph edge can be transferred to matter, and vice versa.  That is, starting from the configuration 
\be
	\begin{array}{c}\mbox{\includegraphics[height=1.5cm]{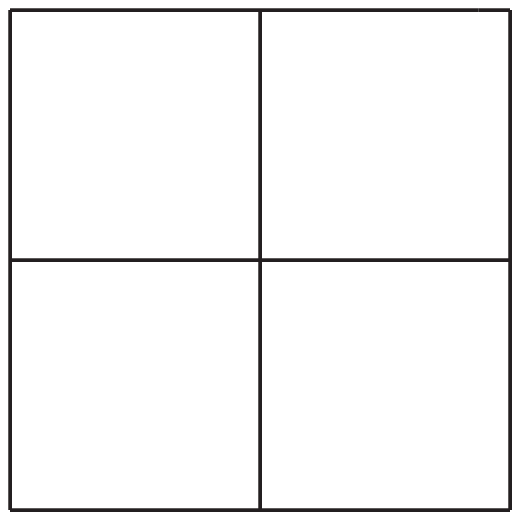}}\end{array}\nonumber
\ee
we can erase a link to create two bosons on the vertices of the erased link (the black dots represent bosons):
\be
	\begin{array}{c}\mbox{\includegraphics[height=1.5cm]{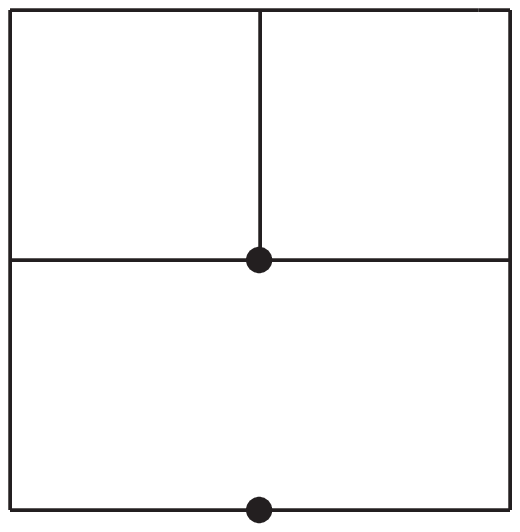}}\end{array}\nonumber
\ee
The above process is reversible, i.e., we can eliminate two bosons and replace them with a link connecting their supporting vertices.  This exchange is done using the term in the Hamiltonian\footnote{
In \cite{HMLCSM}, we introduced the more general Hamiltonian that created $R$ pairs of  bosons for every link destroyed, and vice versa:
\be
H_{\rm ex}=k\sum_{ij}\left(|0\rangle\langle1|_{ij}\left(b^\dagger_i b^\dagger_j\right)^R+
		|1\rangle\langle 0|_{ij}\left( b_ib_j\right)^R\right).
\ee
Here, for simplicity, we take $R=1$.
}
\be
H_{\rm ex}=k\sum_{ij}\left(|0\rangle\langle1|_{ij}\ b^\dagger_i b^\dagger_j+
		|1\rangle\langle 0|_{ij}\ b_ib_j\right).
\label{eq:Hex}
\ee

In addition, the bosons can hop around the lattice {\em but only where a link exists}, i.e., they can hop from $i$ to $j$ only if the link $ij$ is on:
\be
H_{\rm hop}=-t\sum_{ij}|1\rangle\langle 1|_{ij}
\left( b_i^\dagger b_j +b_i b_j^\dagger\right).
\ee
This is an important feature of the model as it means that {\em it is the behavior of the matter that gives the graph the meaning of geometry}.  {An edge between two nodes $i,j$ means that there is a hopping term in the Hamiltonian between $i$ and $j$. Since the meaning of the edges is given by the dynamics of the particles, a pure graph without particles is just a mathematical structure without physical content. To the extend that such a model can capture aspects of gravity,  pure gravity will not be meaningful. }

The combination of $H_{\rm ex}$ and $H_{\rm hop}$ cause the lattice to change, as for example in this sequence:
\be
	\begin{array}{c}\mbox{\includegraphics[height=1.5cm]{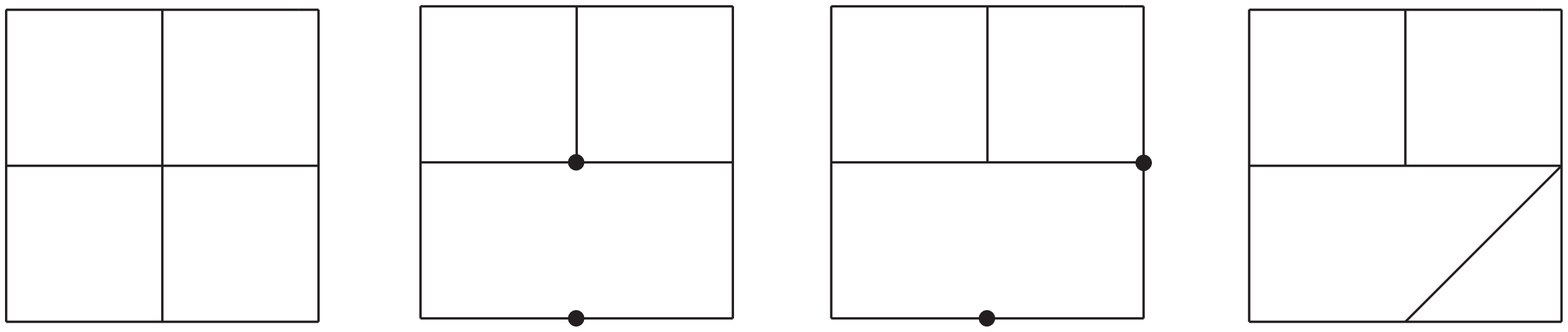}}\end{array}\nonumber
\ee
Note that it is more likely that bosons will be created in more highly connected parts of the lattice, as there will be a higher density of links to be turned into matter, and, similarly, new links are more likely to be created where there is a higher concentration of matter.

With the term $H_{\rm ex}$ as given by equation (\ref{eq:Hex}) above it is possible to destroy to bosons and create a new graph link in their place, no matter how far apart, with respect to the lattice, the bosons are.  The following sequence is an example:
\be\label{sequence}
	\begin{array}{c}\mbox{\includegraphics[height=1.5cm]{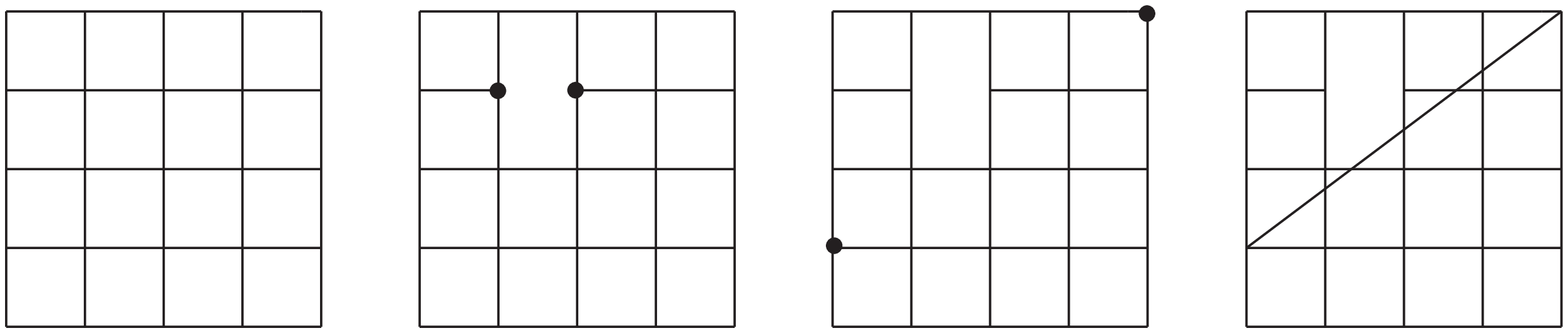}}\end{array}\nonumber
\ee
In \cite{HMLCSM}, we eliminated the possibility of such non-local  connections by allowing boson-edge exchange only if there already is a path of length $L$ connecting the boson sites, where $L$ is some short distance:
\be
H_{\rm ex}=k\sum_{ij}\left(|0\rangle\langle1|_{ij}P^L_{ij}\ b^\dagger_i b^\dagger_j+
		P^L_{ij}|1\rangle\langle 0|_{ij}\ b_ib_j\right).
\label{eq:HexP}
\ee
where 
\be
P^L_{ij}=\sum_{k_1,...,k_{L-1}} P_{ik_1}P_{k_1k_2}\cdots P_{k_{L-1}j};\qquad P_{ij}=|1\rangle\langle 1|_{ij}.\ee
In what follows, we will take $L=2$.  Note that, with the above dynamics the graph cannot disconnect.

The total Hamiltonian for this model then is
\be
H=H_{\rm link}+H_v+H_{\rm ex}+H_{\rm hop}.
\label{eq:H2}
\ee

\subsection{A toy black hole.} 
\label{toyBH}
 As we discussed in the introduction, our working interpretation of the physical content of General Relativity is that geometry tells matter where to go and matter tells geometry how to curve.  Can the reverse be true, i.e. in a system where geometry tells matter where to go and matter tells geometry how to curve, do we see aspects of gravity?

The model we just defined is one where geometry (the graph) determines where matter is allowed to go, while matter influences geometry: a particle at vertex $i$ and one at vertex $j$ can be replaced by a new edge $ij$ where one did not previously exist and so change the distance between $i$ and $j$.  This is not quite the same as General Relativity, but the model does have some features that are reminiscent of gravitational behavior.  We will illustrate this by constructing a toy ``black hole'' in our spin system.  

Consider a state of the system in which the graph is a flat lattice but with a region that is highly connected.  We call the flat region $A$ and the highly connected region $B$:
\be
	\begin{array}{c}\mbox{\includegraphics[height=5cm]{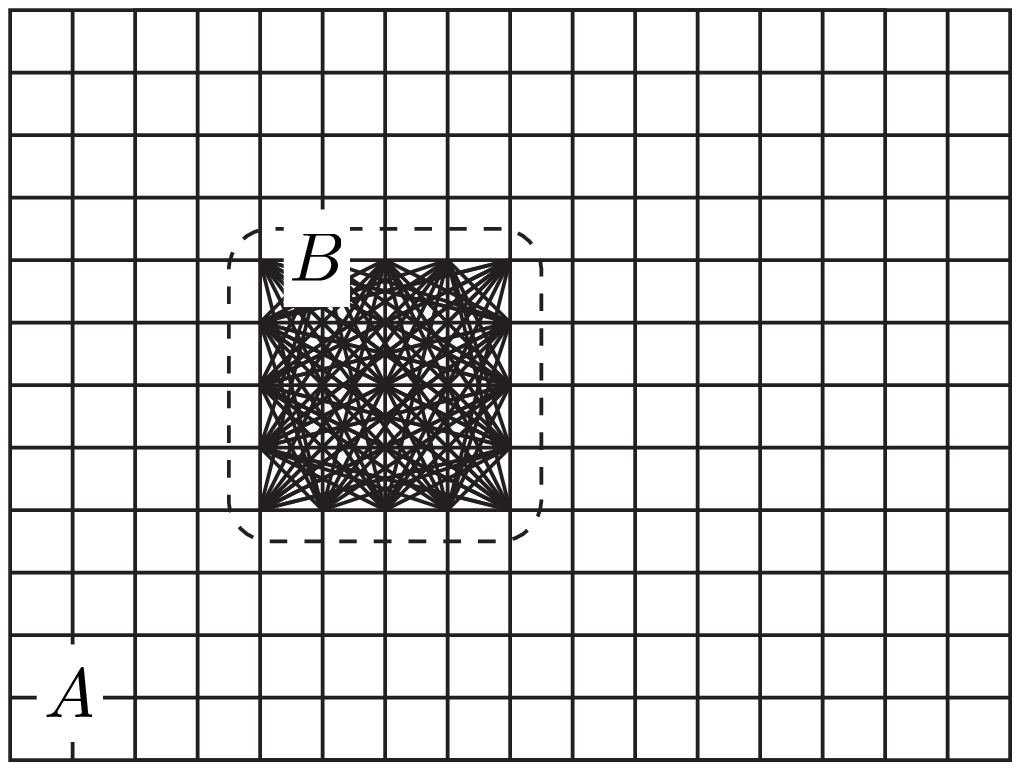}}\end{array}\nonumber
\ee
To make things simple, if $B$ contains $N_B$ vertices, we ask that $B$ is completely connected so that the degree of vertices in $B$ is $\sim N_B$.  

 It is intuitively clear that $B$ will act as a trap: there are many more links to the interior of $B$ than paths leading to the outside $A$.  
A boson that hops its way to the boundary of $B$ is far more likely to take its next step into $B$'s interior and once inside there are many more edges that will keep it inside than edges leading back to $A$. 
 This intuition can be made more quantitative by using the Lieb-Robinson 
 speed of light that we defined and calculated in Section \ref{SectionLR} to find out what happens to a ray of light traveling on this lattice\footnote{
We can add Wen's $U(1)$ Hamiltonian for emergent light \cite{Wen} in eq.\ \ref{eq:H2}.  In the phase where the couplings of the $U(1)$ theory are small with respect to the other couplings in the Hamiltonian, we have electromagnetic waves traveling on the lattice.
}.
Recall that we found that  the LR speed is proportional to the degree of the lattice.  Hence, in region $A$, $c_A\sim 1$, while in region $B$, $c_B\sim N_B$.  Now consider a particle traveling in this graph and crossing the boundary between $A$ and $B$. Snell's law states that the critical angle for total internal reflection is 
\be
\theta_c=\sin^{-1}\frac{c_B}{c_a}=\sin^{-1}N_B, 
\ee
and so the probability of a light ray escaping $B$ is of the order of $N_B^{-1}$.  In the large $N_B$ limit, region $B$ traps light and matter.  

$N_B$ is not infinite and so light and matter can eventually  escape.  Assuming that $A$, even though less dense, is much larger than $B$,  $B$ will evaporate until its edges are of a density comparable to $A$.  The whole process is of course unitary, but the emitted quanta are entangled with the remnant in $B$.  That is, the spectrum of the emitted radiation will be mixed even though the underlying dynamics is unitary.  To the extend that this configuration can be thought of as a toy black hole, it illustrates the resolution of the information loss paradox via matter/geometry entanglement and no black hole singularity.  

Of course, there is no claim here that this is a real black hole.  There are important differences between what we just described and black holes in General Relativity: this is not a spacetime construction; the horizon can be seen by a local observer crossing it.  We really are modeling a black hole using two media of different refractive indices.  Nonetheless, it illustrates that in this model there is an entropic form of attraction: highly connected regions attract bosons in the sense that the particles are more likely to be found at vertices of higher degree.  In addition, there is an analogue of negative heat capacity: bosons are more likely to be created in the highly connected parts of the lattice as there there is a higher density of links that can be turned into matter, and new links are more likely to be created where there is a higher concentration of matter.


\section{Matter/geometry entanglement and thermalization}
\label{entanglement}

One of the advantages of approaching the problem of quantum gravity using
 dynamical lattice by means of spin systems and hopping particles, is that we have at our hand methods of quantum information and many body theory.
One of the features of the model of the previous section  is  the tensor product structure $\mathcal H = \mathcal H_{\rm graph} \otimes \mathcal H_{\rm matter}$ in the state space. It is natural  to ask  what  the role of entanglement is with respect to such bipartition. In other words, how much can matter and space get entangled? This question is relevant in understanding the sense in which matter and space, taken separately, can thermalize. 

\subsection{Thermalization via subsystem dynamics.}
The quantum evolution of the whole system is unitary and is given by
 \be\label{uni}
\rho(t) = U(t) \rho(0) U^\dagger(t),
 \ee
where $U(t) =
e^{-iHt}$. Now, suppose we start with a pure state $\rho(0)$ which is a product state with respect to the bipartition $\mathcal H = \mathcal H_{\rm graph} \otimes \mathcal H_{\rm matter}$.  If the unitary
$U(t)$ acts as an entangling gate, the state $\rho(t)$ will be entangled. The evolution of the matter alone will be given by 
 \be\label{partial}
  \rho_{\rm matter}(t) = \mbox{Tr}_{\rm graph}\rho(t)=
\mbox{Tr}_{\rm graph}\left[ U(t) \rho(0) U^\dagger(t) \right].
\ee
The quantum evolution for the matter only is that of an open system and thus is not unitary. That is, the state $ \rho_{\rm matter}(t)$ can  be mixed. The question then is whether there is enough entanglement to make $ \rho_{\rm matter}(t)$ a thermal state. A first thing to note is that the dimension of the matter Hilbert space is $\dim\mathcal H_{\rm matter} = N\dim \mathcal H _i$ while the dimension of the Hilbert space of the graph is $\dim\mathcal H_{\rm graph} = N^2\dim \mathcal H _e$. So, as long as $\dim \mathcal H _e \simeq \dim \mathcal H _i$, it is possible for the matter to be in a completely mixed state, and therefore have a thermal radiation up to infinite temperature. The details of the model determine what kind of thermal radiation is possible to get.

On the other hand,  the quantum evolution of the graph will also be that of an open quantum system and therefore the graph will not evolve unitarily. In the right regime of the couplings, namely a weak coupling between graph and matter,  an effective thermal behavior for the graph is possible. For this to happen, 
first of all, we need the Hamiltonian for the graph to be non-degenerate. Although $H_{\rm graph}$ is highly degenerate, such degeneracy is easily lifted by random perturbations. Now, let $\sigma_m (E)$ be the density of states given by the Hamiltonian $H_{\rm matter}$. We can treat the matter as a heat bath for the graph with inverse temperature
\be\label{efftemp}
\beta (E) = \frac{d\ln \sigma (E)}{dE}.
\ee
The effective temperature $\beta (E)$ is set by the energy of the initial state of the system $\rho (0)$, which is supposed to be separable $\rho(0) = \rho_{\rm graph}^0 \otimes \rho_{\rm matter}^0$ and such that $\rho_{\rm matter}^0$ has  energy peaked around a certain value $E$.
Let us also suppose that the interaction Hamiltonian $H^{\rm I} = H_{\rm hop} + H_{\rm ex}$ is a perturbation,  describing scattering processes that approximatively conserve the unperturbed energies (weak coupling). Now consider any observable $A$ on the graph. These observables comprise all the information we can extract about geometry in our system. The quantum evolution of the expectation value $A (t)$ is of course given by 
\be
\langle A \rangle (t) = \tr \{ \rho(t) A \otimes {\bf 1}_{\rm matter} \}.
\ee
Under the hypothesis above, it is possible to prove \cite{tasaki} that the long time value of $\langle A \rangle (t) $ is given by

\be
\lim_{t\rightarrow\infty} \langle A \rangle (t)  \simeq \frac{
\tr_{\rm graph}\{ 
A e^{-\beta (E) H_{\rm graph}}
\}
} {
\tr_{\rm graph}\{ 
e^{-\beta (E) H_{\rm graph}}
\}
}.
\ee
This means that, even if the total system is not at the equilibrium and is evolving unitarily, and in particular the particles are not in some thermal equilibrium, nevertheless there exists an effective temperature $\beta (E)$ such that the large time limit of the expectation value of any observable on the graph gives the canonical expectation value at the temperature $\beta (E)$.
This is a rather old idea (see for example, \cite{Pag}), which we hope finds a concrete application in our system.  
 It is the subject of our current research to show that a similar result can be valid also for system Hamiltonians that can produce an interesting geometry.

{ Equilibration in probability.}
A related approach to the issue of equilibration in a closed quantum system is to look at equilibration in probability. 
This approach can be used to study both models presented here. Even if the long time limit of the expectation value of an observable $\widehat{A}_L$ does not exist, they would still spend most of their time very close to their time average $\overline{A}$. Here the subscript $L$ reminds us that the system has a finite (linear) size $L$. To be more precise, consider the time average $\overline{\rho}= \lim_{t\rightarrow\infty} \tau^{-1}\int^\tau_0 \rho(s) ds$, which always exists and is given by the $\rho_0$ totally dephased in the eigenbasis  $\Pi_n = \ket{E_n}\bra{E_n}$ of the Hamiltonian: $\overline{\rho}= \sum_n\Pi_n \rho_0 \Pi_n$.  We suppose the spectrum $\{ E_n\}$ of the Hamiltonian to be non-degenerate. This is not a very strict condition since a random perturbation of the Hamiltonian will lift any degeneracies. 
The time average of the observable $\widehat{A}$ is given by
\be\label{avg}
\overline{A (t)} = \lim_{\tau\rightarrow\infty} \frac{1}{\tau} \int^\tau_0 A(t) dt = \tr \left(\sum_n\Pi_n\rho_0\Pi_n\widehat{A}\right)=\tr \{\overline{\rho} \widehat{A}\}.
\ee
Equilibration in probability means then that
\be\label{eqpr}
\lim_{L\rightarrow\infty} Pr \{ |A_L (t) -\overline{A (t)}|\ge \epsilon \} =0.
\ee
Equilibration in probability is a typical feature of quantum many-body systems away from equilibrium. The equilibrium expectation values $\overline{A}$ do not need to be those given by the canonical distribution and therefore this is a much weaker feature.

\subsection{Thermalization in our model}

\begin{figure}
  \includegraphics[width=10cm]{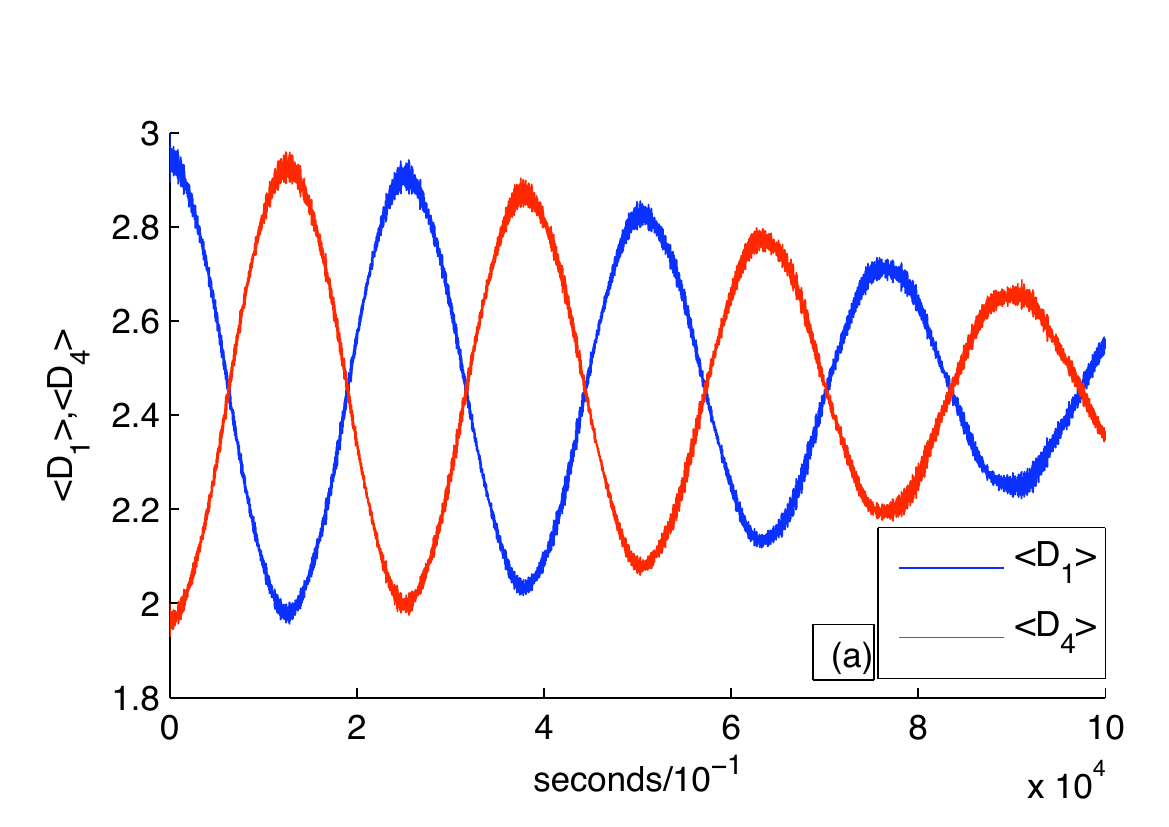}\\
  \caption{The time evolution for the expectation value of the degree number observables $D_1, D_4$ at two nodes $1,4$ for the system in the Mott insulating phase. The two values are oscillating but equilibrating towards a common value independent on the initial state.}
  \label{eqprob}
\end{figure}

\begin{figure}
  \includegraphics[width=10cm]{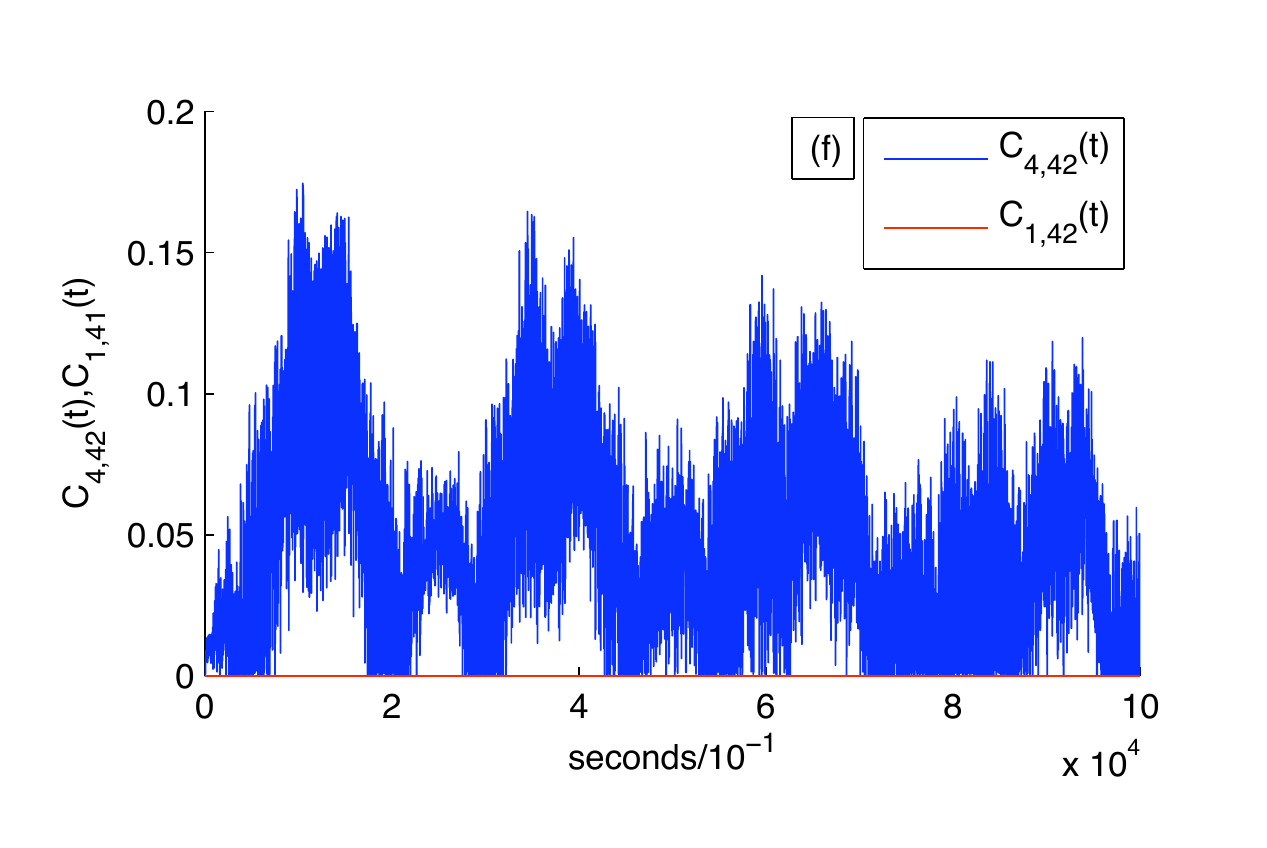}\\
  \caption{Entanglement dynamics in the model. Here we plot the concurrence between the number of particles $n_4$ at the node $4$ and the insisting edge $24$ (blue line). We see a non trivial evolution which shows equilibration at large times. Red line shows the same quantity for the node $1$, showing that an edge only entangles with the nodes it comes from and therefore $C_{1,42}=0$ during the whole evolution.}
  \label{entdyn}
\end{figure}

In order to study equilibration and thermalization in this model, we need to resort to numerical simulation. Even to simulate the small graph with four nodes we need to consider
the Hamiltonian Eq.(\ref{eq:H2}) for hard core bosons so that the local Hilbert space of a vertex is just two-dimensional. in the regime where the hopping and exchange term are a perturbation of the potential energy terms $H_{\rm link}$ and $H_{\rm hop}$. This is the regime that in condensed matter physics is called Mott insulator when referring to the Hubbard model: particles are localized and there is no transport. Eigenstates of the model are then approximatively product states of the edge/particle configurations, which allows for a direct evaluation of Eq.(\ref{avg}). The entanglement dynamics and the signs of thermalization for the model Eq.(\ref{eq:H2}) with just $4$ vertices have been studied numerically in \cite{HMLCSM}, showing that thermalization occurs in the Mott insulating phase of the model. As it can be seen in Fig.\ref{eqprob}, the expectation values of the degree operators evolve towards a typical value. In the Mott insulating phase thus, we have at least equilibration in probability. Moreover, we see that there is a non-trivial entanglement dynamics. In Fig.\ref{entdyn} we show the time evolution of the concurrence between a vertex and an edge attached to it. Also this entanglement is slowly damping, showing a sign of equilibration.

\begin{figure}
  \includegraphics[width=12cm]{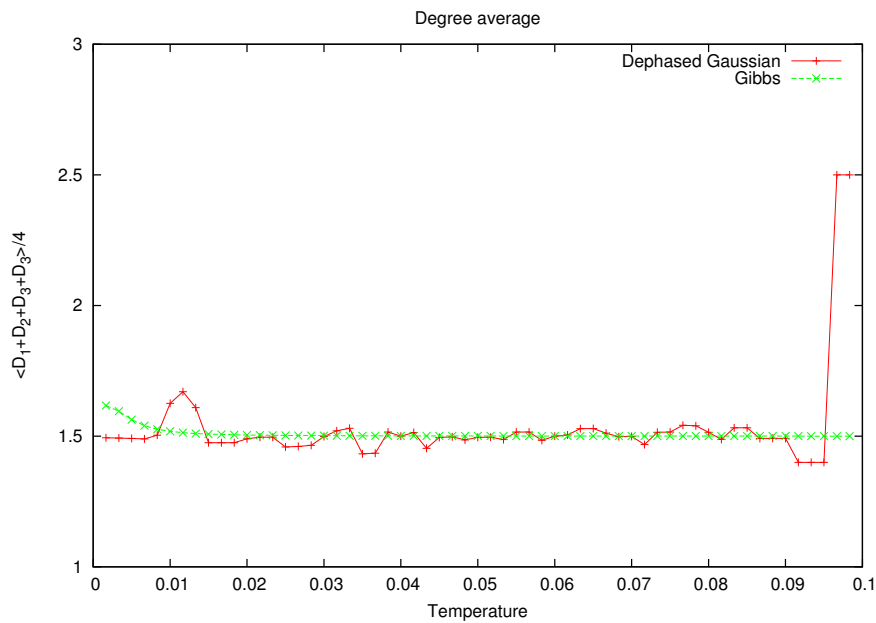}\\
  \caption{The long time limit expectation value of $D$ is the red curve with the crosses,  while its thermal value is represented by the green curve with the x. For a large range of temperatures the two expectation values are very similar. }
  \label{Gibbs}
\end{figure}

As we have seen, equilibration in probability means that the long time expectation values of observables are given by Eq.(\ref{avg}). We would like to know if in our model equilibration in probability gives \textit{thermal} expectation values for the relevant observables. Let us consider an initial state $\rho_0 = |\psi_0\rangle\langle \psi_0|$ that is very peaked around a certain energy $E_0$. We choose a properly normalized initial state $|\psi_0\rangle = \alpha\sum_n \exp ((E_n-E_0)^2/2\sigma) |E_n\rangle $.  Then we can compute, say, the long time expectation value $\overline{D}$ of the observable $D$ that measure the average degree of the nodes in the graph. By Eq.(\ref{avg}) one can see that this value is a function of the initial state $\rho_0$ and therefore of the effective temperature $\beta (E_0)$. Then we can consider the Gibbs state at the same temperature and compute the thermal expectation value of $D$. The two curves are shown in Fig.\ref{Gibbs}. For a large range of temperatures the two expectation values are very similar. Therefore the typical, equilibration value of such observable corresponds to the thermal one at the effective temperature $\beta (E_0)$. A systematic study of the thermalization properties of this kind of model is the subject of current research \cite{th}.

\section{Outlook and Conclusions}
\label{Future}

The research program we have outlined in this article presents numerous new possibilities and new issues.  In this final section, we will sketch out some of those, for the central issues of time and gravity, and quantum mechanics.

{\em  Time and Lorentz invariance.} 
It is natural for a reasearcher in quantum gravity with a background in relativity to object to the kind of approach we are advocating here.  The spin system has a fundamental Hamiltonian, that is, we are using non-relativistic quantum mechanics.  How can we possibly reconcile this with time in relativity?  
In fact, there are a number of different possibilities, some of which we will sketch out here: 

\noindent 1.  Effective finite lightcones from non-relativistic quantum mechanics.  
The physics of the Lieb-Robinson speed that we discussed in section \ref{SectionLR} provides a way to have both non-relativistic quantum mechanics and finite lightcones.  These become reconciled simply because, even though there {\em is} a signal outside the lightcone, it is exponentially suppressed.  
Note that the Lieb-Robinson lightcone is an effective lightcone  but cannot be called emergent as it is still at the level of the spin system.  Whether, combined with the string network mechanism and the emergent Maxwell equations, we not only have finite lightcones but also have emergent Minkowksi spacetime needs to be investigated.  

\noindent 2.  Microscopic vs emergent internal time.
At least naively, condensed matter-type approaches to quantum gravity must break Lorentz invariance because of the presence of a lattice and the associated fundamental discreteness and preferred frame. Recently there has been sub- stantial attention to Lorentz invariance violations and there are tight constraints on certain types of violations, with more data coming in in the near future. Nonetheless, there are still numerous possibilities for Lorentz invariance violations that are not so constrained \cite{LIV}.
Perhaps a little less naively, the relevant question is what is the physics seen by an observer that lives inside the system and who is at low energy and hence has no access to the Planckian lattice. For example, in analog gravity \cite{analog}, the external metric is flat Newtonian, while the internal is Lorentzian. 

An important issue to emphasize is that the models of \cite{KoMaSe} and \cite{HMLCSM}, as well as condensed matter approaches to quantum gravity in general, assume the existence of a notion of time and of time evolution as given by a Hamiltonian, as opposed to the constrained evolution of canonical pure gravity. It is a general question for all condensed matter approaches to quantum gravity whether such evolution is consistent with the diffeomorphism invariance of general relativity. While it is not possible to settle this question without first finding out whether the condensed matter microscopic system has a low energy phase which is general relativity, we can analyze the problem further, following the direction outlined in \cite{FQXi}. In general, there are two possible notions of time: the time related to the $g_{00}$ component of the metric describing the geometry at low energy and the time parameter in the fundamental microscopic Hamiltonian. Let us call the first geometric time and the second fundamental time. In our context, it is clear that the geometric time will only appear at low energy, when geometry appears. The problem of the
emergence of geometric time is the same as the problem of the emergence of space, of geometry. The constrained evolution of general relativity  refers to geometric time. By making the geometry not fundamental, we are able to make a distinction between the geometric and the fundamental time, which opens up the possibility that, while the geometric time is a symmetry, the fundamental time is real. It is important to note that the relation between geometric and fundamental time is non-trivial in the systems we are studying and that the existence of a fundamental time does not necessarily imply a preferred geometric time. We also note that, in the presence of matter in general relativity, a proper time can be identified. In addition, the system studied in \cite{HMLCSM} has matter and in that sense it is perhaps more natural that it also has a straightforward notion of time.

\noindent 3.  Lorentz invariance as a property of the matter.  
Another point, relevant to these kinds of approaches to quantum gravity, is that
Lorentz invariance can be viewed as a symmetry of the geometry--the physics of Minkowski spacetime--or as a symmetry of the matter.  The latter is how Lorentz viewed his transformations: when a charged particle moves, its field lines get deformed (see, for example, \cite{Bell,OD}).  Viewing the Lorentz transformations as a property of the spacetime amounts to {postulating} that there is a maximum speed and that all massless particles travel at that speed.  In Einstein's theory, there is no reason why this is so, but introducing the postulate by introducing the Minkowski spacetime is a very elegant way to impose universality of the speed of light.  Emergent gravity approaches do not have this option.  Instead, it is a common feature, and perhaps the most important challenge to such approaches, that different particle species can travel at different speeds.  

Obtaining relativistic spacetime out of a microscopic quantum mechanical system amounts to showing that diffeomorphisms is an emergent symmetry.  It makes sense to start with Lorentz invariance as an emergent symmetry, as an intermediate step, especially given the recent attention that has been given to 
 possible observational signatures of violations of Lorentz invariance \cite{LIV} and the possibility that Lorentz invariance mat be emergent \cite{JacLI}.  Since research in quantum gravity has for long suffered from lack of access to observations, it is a very attractive feature of emergent gravity approaches that observations can prove them wrong, and in the near future.

{\em Quantum effects of dynamical geometry.}
It is common to expect that a quantum theory of gravity will, at some level, be described by a quantum superposition of geometries.  This superposition presents many mathematical and conceptual problems.  A very important one is the emergence of our classical world:  why do we not see superpositions of geometry in every day life?  This is in effect the measurement problem in a cosmological context, one of the topics of study in quantum cosmology.  For us, the geometry is given by quantum states describing graph configurations, and we have indicated a mechanism for the emergence of metastable geometries from unitary evolution.   As long as the metastable states do not present macroscopic entanglement, the coarse grained observables would behave as if there is no superposition. But why we do not have such macroscopic entanglement? This is an open problem but our approach makes it attackable. If the state we consider is the ground state of a gapped local Hamiltonian, all correlations functions decay exponentially and the total amount of entanglement is very low. The behavior of the system away from equilibrium is more complicated, but lately there has been a flourishing of results about the time evolution of quantum many body systems after a quantum quench \cite{quench}, and our models can be studied in the same way. 

{
At very short scales, however, it is possible to have an equal superposition of different geometries. Time evolution can for instance produce a state that locally looks like a superposition of the first and fourth state of Eq.(\ref{sequence}). This superposition is tensored with the state of the square lattice everywhere else. Imagine we perform an experiment on the time of arrival of particles to the upper right corner. When the system collapses in the first branch, we will have a certain time of arrival $t_1$. On the other hand, when the system collapses in the second branch, a particle can jump directly from the lower left corner to the upper right corner, yielding an arrival time $t_2 < t_1$. On a large scale, in every branch there is some of this shortcuts and they average out. But at short scale, one can observe the superposition of the state with and without a shortcut. There is a second important quantum effect to consider. Since matter and geometry can be entangled, the evolution of the matter alone cannot be unitary. The evolution of the matter alone must be described in the setting of an open quantum system, and there is decoherence towards the spatial degrees of freedom. Therefore our model implies (at low couplings) a tiny deviation for unitarity. Moreover, the decoherence rate depends on the entanglement and thus on the average degree of the graph. Since high connected graphs correspond to strong curvature, we expect to only observe decoherence in presence of strong gravitational effects. The entanglement between matter and geometry has also consequences for the black hole information paradox. If we consider the matter escaping from our toy black hole, we will observe a mixed state, even if we have started with a pure state. Because of the open system scenario, the black hole behaves like a quantum channel that increases the entropy. In other words, the matter coming out is thermal because it is entangled with what is inside the black hole, both edges and particles. When all the matter has escaped, there is no matter to be entangled with anymore. This is what makes the problem in the usual treatments of the black hole information paradox: we have thermal radiation coming from the entanglement of matter with nothing! But in our scenario, the disappearance of the black hole only means that the graph rearranges in a way such that the region that hosted the black hole becomes homogenous with the surrounding space. The radiation is still entangled with all the spatial degrees of freedom in that region. The black hole can disappear but the entanglement stays. 

To conclude, we are only at the beginning. We need to give a detailed description of the open quantum system to the end of making predictions about the decoherence in presence of strong fields. We also need to exploit entanglement theory in quantum many body systems to give a precise description of the entanglement in these models, and proving that macroscopic entanglement is negligible. 

Even when we have accomplished all this, we still do not have gravity! The way the graph curves in presence of matter in our model has nothing to do with what gravity does. The Hamiltonian for the graph and its interaction with the matter is too simple. The gravity model will be very complicated, with many messy terms. We think that the fundamental theory is messy, and the emergent theory is simple and elegant. So we do not go to smaller scales to \textit{find} the simple fundamental theory, but to \textit{explain} the beauty of the emergent one.}\\


\section*{Acknowledgments}

The authors are indebted to Abhay Ashtekar, Francesco Caravelli, Florian Conrady, Bei-Lok Hu, John Preskill, and Lee Smolin
useful advice and comments. 
We are grateful to the Santa Fe Institute where this work was carried out and its members for their wonderful hospitality. 

This work was supported by the Alexander von Humboldt Foundation and NSERC grant RGPIN-312738-2007.  
Research at Perimeter Institute is supported by the Government of Canada through Industry Canada and by the Province of Ontario
through the Ministry of Research \& Innovation.


\end{document}